\documentclass[twocolumn,showpacs,preprintnumbers,amsmath,amssymb,prl,superscriptaddress]{revtex4}
\usepackage{mathrsfs}
\usepackage{graphicx}
\usepackage{dcolumn}
\usepackage{bm}
\usepackage{amsmath}
\usepackage{amsfonts}
\usepackage{color}

\begin{document}
\title{Wave Function and Emergent SU(2) Symmetry in $\nu_T=1$ Quantum Hall Bilayer}
\author{Biao Lian}
\affiliation{Department of Physics, McCullough Building, Stanford University, Stanford, California 94305-4045, USA}
\affiliation{Princeton Center for Theoretical Science, Princeton University, Princeton, New Jersey 08544-0001, USA}
\author{Shou-Cheng Zhang}
\affiliation{Department of Physics, McCullough Building, Stanford University, Stanford, California 94305-4045, USA}

\begin{abstract}
We propose a trial wave function for the quantum Hall bilayer system of total filling factor $\nu_T=1$ at a layer distance $d$ to magnetic length $\ell$ ratio $d/\ell=\kappa_{c1}\approx1.1$, where the lowest charged excitation is known to have a level crossing. The wave function has two-particle correlations which fit well with those in previous numerical studies, and can be viewed as a Bose-Einstein condensate of free excitons formed by composite bosons and anti-composite bosons in different layers. We show the free nature of these excitons indicating an emergent SU(2) symmetry for the composite bosons at $d/\ell=\kappa_{c1}$, which leads to the level crossing in low-lying charged excitations. We further show the overlap between the trial wave function and the ground state of a small size exact diagonalization is peaked near $d/\ell=\kappa_{c1}$, which supports our theory.
\end{abstract}

\date{\today}


\maketitle


The two-dimensional (2D) electron quantum Hall bilayer exhibits a rich physics \cite{Suen1992,Eisenstein1992}. When the system has a total filling factor $\nu_T=1$ and a negligible interlayer hopping, an exciton superfluid phase arises at small layer distances $d$ \cite{Murphy1994}, which exhibits a perfect Coulomb drag effect \cite{Gramila1991,Kellogg2002,Kellogg2004,Tutuc2004} and a greatly enhanced interlayer tunneling \cite{Spielman2000}. The electrons and holes in the lowest Landau levels (LLLs) of different layers are bounded into excitons, and form an exciton superfluid which has a U($1$) symmetry breaking and a charge gap \cite{Wen1992b,Wen1993,Ezawa1993,Yang1994,MacDonald1994,Moon1995,Fogler2001,Yang2001,Eisenstein2004}. As the layer distance $d$ over the magnetic length $\ell$ exceeds a critical value $d/\ell>\kappa_{c2}\approx 1.8$, the system enters a compressible phase without superfluidity if the bilayer is symmetric \cite{Murphy1994}, which is equivalent to two copies of the $\nu=1/2$ composite fermi liquid (CFL) as $d\rightarrow\infty$ \cite{Halperin1993,Son2015,Kim2001}.

Constructing trial wave functions for quantum Hall systems proves to be a powerful and successful method \cite{Klitzing1980,Tsui1982,Suen1992,Eisenstein1992,Laughlin1983,Halperin1983,Haldane1988,Read1996,Zhang1992,Wen1992a}. In the limit $d\rightarrow0$, the ground state of the $\nu_T=1$ bilayer is known to be the Halperin ($111$) state \cite{Wen1992b,Yang2001,Halperin1983}.
Meanwhile, the ground state of the system at intermediate layer distances $0< d/\ell< \kappa_{c2}$ is still unsettled
\cite{Cote1992,Bonesteel1996,Stern2002,Veillette2002,Alicea2009,Cipri2014,Doretto2006,Doretto2012,Moller2008,Simon2003,Shibata2006,Zhu2017}, which is a major obstacle in understanding the transition from exciton superfluid to CFL. Numerical calculations are employed to reveal the nature of the ground state \cite{Nomura2002,Moller2008,Simon2003,Shibata2006,Zhu2017,Park2004,Schliemann2001}, and a charge gap closing is indeed observed at $d/\ell\approx 1.8$. In particular, the calculations of both density matrix renormalization group (DMRG) and exact diagonalization (ED) have identified a level crossing between the first and second charged excitations at a layer distance $d/\ell=\kappa_{c1}\approx 1.1$ \cite{Shibata2006,Zhu2017}, which is not yet well understood.

In this letter, we propose a trial wave function for the exciton superfluid of the $\nu_T=1$ bilayer at the level crossing layer distance $d/\ell=\kappa_{c1}$, and show its two-particle correlations fit well with the previous DMRG results \cite{Shibata2006}. The wave function can be viewed as a Bose-Einstein condensate (BEC) of free excitons formed by composite bosons (CBs) and anti-CBs in different layers, based on which we argue there is an emergent SU(2) symmetry for the CBs at $d/\ell=\kappa_{c1}$ that ensures the level crossing. In a crude estimation we obtain $\kappa_{c1}\approx \ln 4\approx1.4$, which is comparable to the numerical result $\kappa_{c1}=1.1$. Lastly, we show the trial wave function has a high overlap with ground state of a small size ED calculation at $d/\ell=\kappa_{c1}$.

we shall take a simplification that all the electrons in the $\nu_T=1$ bilayer are in the LLL of each layer, which is legitimate in the situation $e^2/\epsilon\ell\lesssim \hbar\omega_c$ \cite{Laughlin1983} as is true in the experiments \cite{Murphy1994,Gramila1991,Kellogg2002,Kellogg2004,Tutuc2004,Spielman2000}. Here $e$ is the electron charge, $\epsilon$ is the dielectric function, $\ell=\sqrt{\hbar c/eB}$ is the magnetic length, $\omega_c=eB/m_ec$ is the cyclotron frequency, $B$ is the magnetic field, and $m_e$ is the effective electron mass. We also assume the interlayer hopping is zero. Since the LLL has no kinetic energy, the energy of the system is solely determined by the Coulomb interactions, which take the form $V_{11}(q)=V_{22}(q)=2\pi e^2e^{-q^2\ell^2/2}/\epsilon q$ and $V_{12}(q)=2\pi e^2 e^{-qd-q^2\ell^2/2}/\epsilon q$ in the Fourier space when projected into the LLL \cite{Yang2001,Shibata2006,Zhu2017}, where $q$ is the transferred momentum, and  $V_{ij}(q)$ is the interaction between two electrons in layers $i$ and $j$. It is conventional to define layers $1$ and $2$ as pseudospins $s_z=+1/2$ and $s_z=-1/2$, respectively, so the Hamiltonian of the system has a U($1$) pseudospin rotational symmetry about the $z$ axis.

We first briefly review the ground state of $\nu_T=\nu_1+\nu_2=1$ bilayer in the limit $d\rightarrow 0$, namely, the Halperin (111) state:
\begin{equation}
\Psi_{111}=\mu(z,w)\prod_{i<j}^{N}(z_i-z_j)\prod_{k<l}^{M}(w_k-w_l)\prod_{i,k}^{N,M}(z_i-w_k),
\end{equation}
where $z_i=x_i^{(1)}+iy_i^{(1)}$ and $w_k=x_i^{(2)}+iy_i^{(2)}$ are the complex coordinates of the $i$-th of the $N$ electrons in layer 1 and the $k$-th of the $M$ electrons in layer 2, respectively, and $\mu(z,w)=\prod_i e^{-|z_i|^2/4\ell^2}\prod_k e^{-|w_k|^2/4\ell^2}$ is the Landau level Gaussian factor in the symmetric gauge. The total number of electrons $N+M$ is equal to the Landau level degeneracy, while $\nu_1=\frac{N}{N+M}$ and $\nu_2=\frac{M}{N+M}$ are the filling factors of layers $1$ and $2$, respectively. Note that when $d\rightarrow0$, the Coulomb interactions $V_{11}(q)=V_{12}(q)=V_{22}(q)$ become independent of the layer indices, so the pseudospin rotational symmetry of the system is enhanced to SU(2). As a result, the ground state should reduce to a $\nu=1$ monolayer integer quantum Hall (IQH) state if the layer indices are omitted, which indeed holds for state $\Psi_{111}$.

Unlike most bilayer quantum Hall states, the filling factor difference $\nu_1-\nu_2$ in $\Psi_{111}$ is not fixed, which also holds in the entire exciton superfluid phase. \cite{Shibata2006,Zhu2017}. This leads to a U(1) symmetry breaking order parameter $\Delta(\mathbf{r})=\langle c_1^\dag(\mathbf{r}) c_2(\mathbf{r})\rangle\neq0$ and a charge neutral Goldstone mode, where $c_j(\mathbf{r})$ is the electron annihilation operator at position $\mathbf{r}$ in layer $j$ \cite{Wen1992b,Moon1995,suppl}. Note that $\Delta(\mathbf{r})$ can be viewed as the pairing amplitude of an electron and a hole in different layers, so it marks the occurrence of an exciton superfluid. In a symmetric bilayer with $\nu_1=\nu_2=1/2$, the pseudospin is polarized in the $x$-$y$ plane at an angle $\arg \Delta(\mathbf{r})$ from the $x$ axis.

\begin{figure}[tbp]
\begin{center}
\includegraphics[width=3.4in]{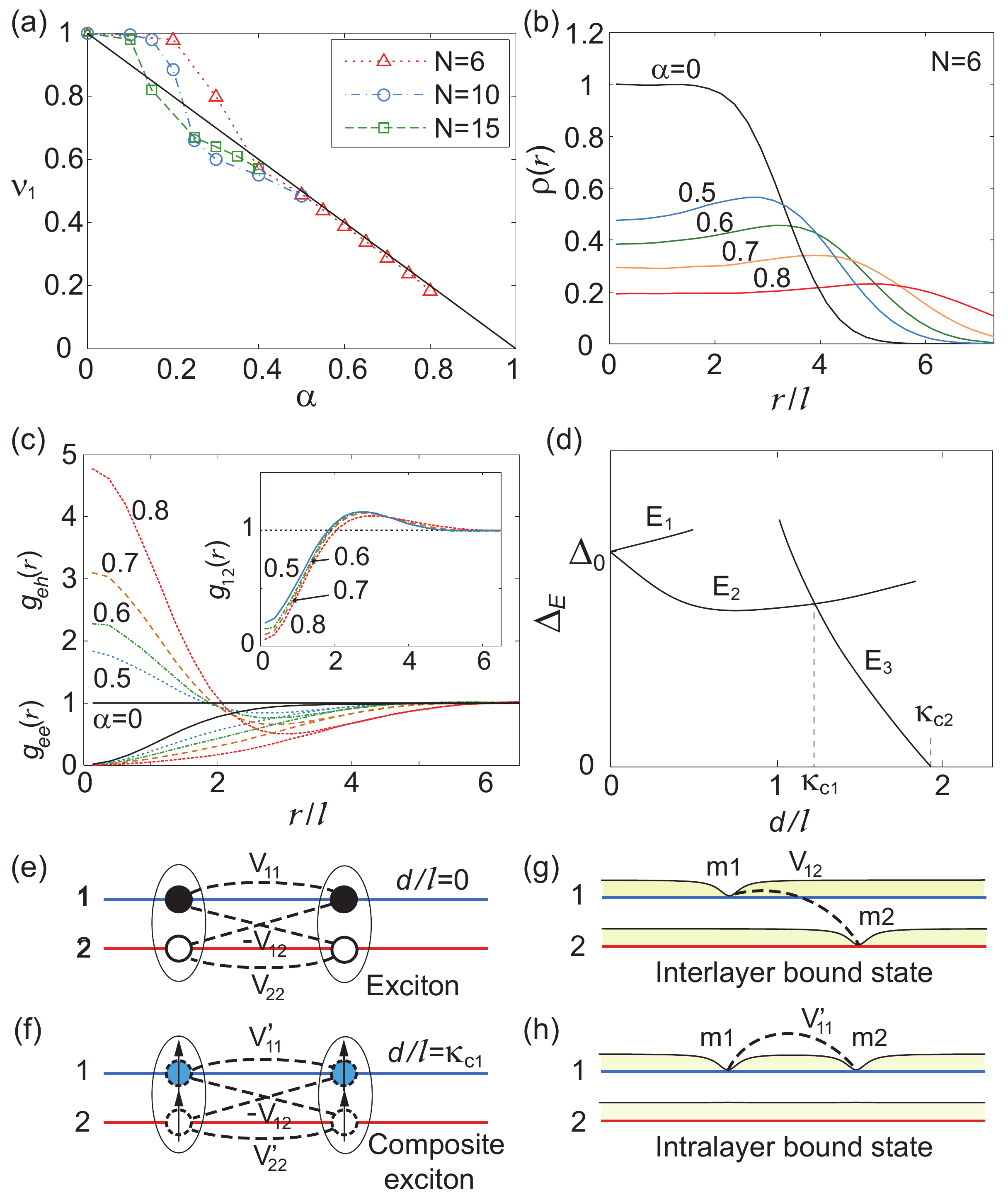}
\end{center}
\caption{(color online) (a) The filling factor $\nu_1$ of $\Psi_{1,\alpha}$ as a function of $\alpha$ for size $N=6,10$ and $15$ from MCMC, in good agreement with $\nu_1=1-\alpha$. (b) The density profile $\rho(r)$ of $\Psi_{1,\alpha}$ for different $\alpha$ and $N=6$. (c) The two-particle correlations $g_{ee}(r)$ and $g_{eh}(r)$ of $\Psi_{1,\alpha}$ for different $\alpha$, and the interlayer correlation $g_{12}(r)$ plotted in the inset. (d) The low-lying charged excitation spectrum expected in our theory, where a level crossing occurs at both $d=0$ and $d/\ell=\kappa_{c1}$. (e)-(f) Interaction between two excitons formed by electrons and holes (e) or CBs and anti-CBs (f). (g)-(h) Illustration of the interlayer and intralayer bound states of two CB merons, which we expect to be excitations $E_2$ and $E_3$ in (d), respectively.}
\label{Fig1}
\end{figure}

The exciton superfluidity of state $\Psi_{111}$ can be seen more clearly via a particle-hole transformation in the LLL of layer $2$ \cite{Girvin1984a,Girvin1984b}, after which the state $\Psi_{111}$ becomes a wave function of $N$ electrons in layer $1$ and $N$ holes in layer 2 as shown in Ref. \cite{Yang2001}:
\begin{equation}\label{Psi0}
\Psi_{0}=\mu(z,w)\sum_{\sigma}\mbox{sgn}(\sigma)\prod_{i=1}^Ne^{z_iw_{\sigma_i}^*/2\ell^2}=\det M_{ij}
\end{equation}
plus a full electron LLL in layer $2$, where $\mu(z,w)=\prod_{i=1}^N e^{-(|z_i|^2+|w_i|^2)/4\ell^2}$ is again the Gaussian factor except that $w_i$ now are the coordinates of holes, $\sigma$ is the permutation of $1$ through $N$ with $\mbox{sgn}(\sigma)$ being its sign, and $\det M_{ij}$ is the determinant of the $N$ by $N$ matrix with elements $M_{ij}=e^{-(|z_i|^2-2z_iw_{j}^*+|w_j|^2)/4\ell^2}$. Both the electrons and holes in such a wave function are in the LLLs at a filling factor $\nu_1$. Note that $|M_{ij}|^2=e^{-|z_i-w_j|^2/2\ell^2}$, so one can view each $M_{ij}$ as a bound state of the $i$-th electron and $j$-th hole, i.e., an exciton wave function. Therefore, $\Psi_0$ can be understood as a BEC of $N$ \emph{free} excitons (similar to the Slater determinant state of $N$ free fermions). The free nature of the excitons is exactly due to the enhanced SU(2) symmetry as $d\rightarrow 0$. As shown in Fig. \ref{Fig1}(e), the Coulomb interaction between two excitons is approximately
\begin{equation}
V_{E}(q)=V_{11}+V_{22}-2V_{12}=2V_{11}(q)(1-e^{-qd}),
\end{equation}
so the exciton interaction $V_{E}(q)$ vanishes as $d\rightarrow0$ \cite{Yang2001}.

At intermediate layer distances $0<d/\ell<\kappa_{c2}$, the excitons become interacting. Since the LLLs have no kinetic energy, the exciton superfluid is strongly correlated and barely understood. A prominent feature revealed by numerical studies is a level crossing between the lowest two charged excited states at $d/\ell=\kappa_{c1}\approx1.1$ for $\nu_1=\nu_2=1/2$ \cite{Shibata2006,Zhu2017}. Besides, unlike $\Psi_{111}$ where the overlap probability of interlayer electrons is zero, the ground state at $d>0$ is shown to have a nonzero interlayer overlap probability \cite{Shibata2006}, which enlarges the intralayer electron spacing and lowers the total energy when $V_{11}(q)>V_{12}(q)$.

We propose here a wave function for $N$ electrons and $N$ holes in layers $1$ and $2$ respectively at certain intermediate layer distances $0<d/\ell<\kappa_{c2}$:
\begin{equation}\label{Psi1}
\begin{split}
\Psi_{1,\alpha}&=\mu(z,w)\prod_{i<j}^N(z_i-z_j)(w_i^*-w_j^*) \sum_{\sigma}\prod_{i=1}^Ne^{\alpha z_iw_{\sigma_i}^*/2\ell^2}\\
&=\prod_{i<j}^N(z_i-z_j)(w_i^*-w_j^*)\mbox{perm}\ M_{ij}(\alpha)\ ,
\end{split}
\end{equation}
where $\alpha$ is a real parameter satisfying $0\le\alpha\le1$, and $\mbox{perm}\ M_{ij}(\alpha)$ is the permanent \cite{perm} of a $N$ by $N$ matrix $M(\alpha)$ with elements $M_{ij}(\alpha)=e^{-(|z_i|^2-2\alpha z_iw_{j}^*+|w_j|^2)/4\ell^2}$.
In particular, we shall show that the state $\Psi_{1,1/2}$ with $\alpha=1/2$ is a good trial wave function for the exciton superfluid in the symmetric $\nu_T=1$ bilayer at $d/\ell=\kappa_{c1}$, and gives an explanation for the level crossing.

The electron (hole) filling factor $\nu_1$ of the wave function $\Psi_{1,\alpha}$ is controlled by the parameter $\alpha$. To see this, we first note the matrix element $M_{ij}(\alpha)$ can be rewritten as $M_{ij}(\alpha)=e^{-[(1-\alpha)(|z_i|^2+|w_j|^2) +\alpha|z_i-w_j|^2]/4\ell^2+i\phi_{ij}}$, where $\phi_{ij}=\alpha(z_iw_{j}^*-z_i^*w_{j})/4i\ell^2$ is real. Ignoring the phase factor $e^{i\phi_{ij}}$ in $M_{ij}(\alpha)$, one can view the translationally invariant part $e^{-\alpha|z_i-w_j|^2/4\ell^2}$ as an exciton bound state, while regard the other part $e^{-(1-\alpha)(|z_i|^2+|w_j|^2)/4\ell^2}$ as the Gaussian factor for a residual magnetic field $(1-\alpha)B$ felt by the electron and the hole. With a Jastrow factor of power $1$ in $\Psi_{1,\alpha}$, one would expect the electrons (holes) to have the same density as that of a $\nu=1$ IQH state in a reduced magnetic field $(1-\alpha)B$, yielding $\nu_1=1-\alpha$. This is verified by our small size Markov chain Monte Carlo (MCMC) calculations \cite{suppl} for state $\Psi_{1,\alpha}$. As shown in Fig. \ref{Fig1}(a), the filling factor for $N=6,10$ and $15$ fits very well with $\nu_1=1-\alpha$ as the system size $N$ increases. Fig. \ref{Fig1}(b) shows the electron (hole) density
$\rho(r)=\langle c_1^\dag(\mathbf{r})c_1(\mathbf{r})\rangle=\langle c_2(\mathbf{r})c_2^\dag(\mathbf{r})\rangle$ as a function of the radius $r=|\mathbf{r}|$ in units of the fully occupied Landau level density for $N=6$ and different $\alpha$, which has a flat droplet shape with an overshoot near the edge similar to that of the Laughlin states \cite{Datta1996,Can2014}.

\begin{figure}[tbp]
\begin{center}
\includegraphics[width=3.4in]{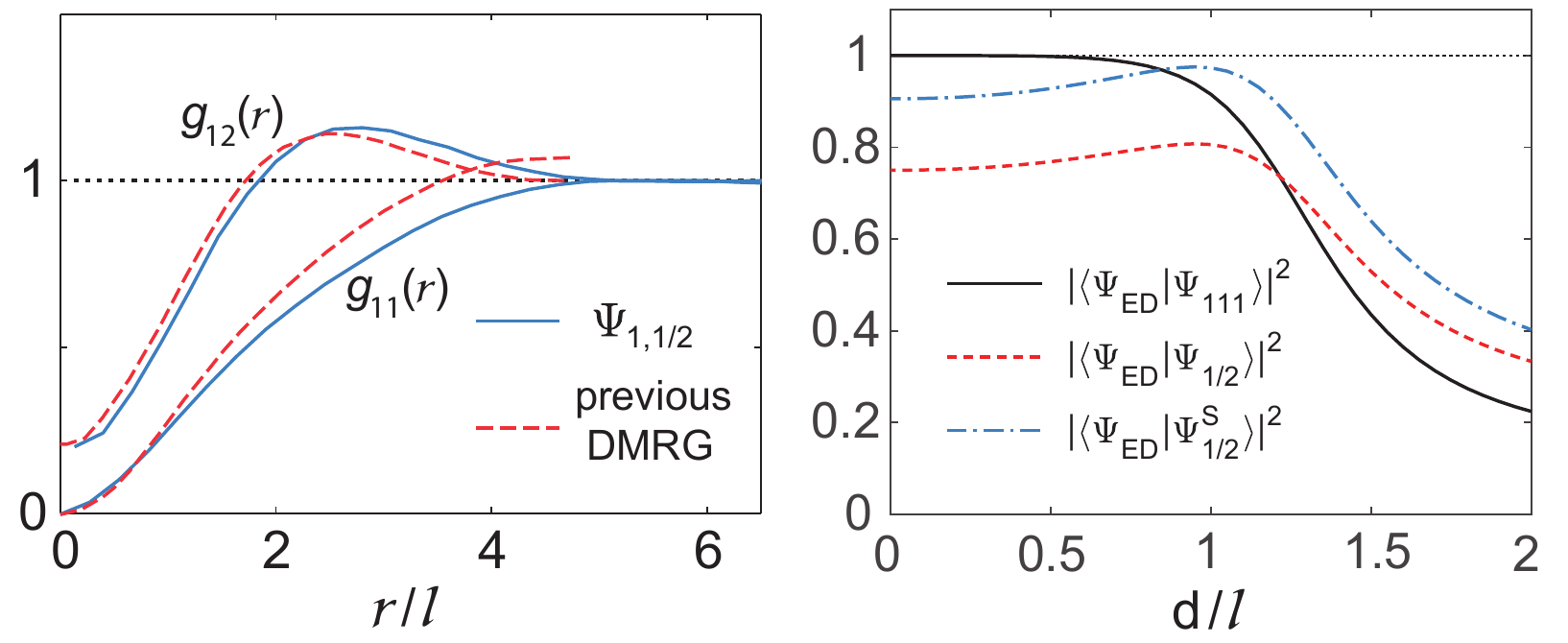}
\end{center}
\caption{(color online) (a) Two-particle correlations $g_{11}(r)$ and $g_{12}(r)$ of state $\Psi_{1,1/2}$ (solid blue lines) and those at $d/\ell=\kappa_{c1}$ reproduced from DMRG calculations in Ref. \cite{Shibata2006} (dashed red lines). (b) Overlap of the ground state $\Psi_{ED}$ from ED calculations for $N=2$ electrons per layer with trial wavefunctions $\Psi_{111}$, $\Psi_{1/2}$ and $\Psi_{1/2}^S$, respectively, as a function of $d/\ell$. }
\label{Fig2}
\end{figure}

The two-particle correlations of $\Psi_{1,\alpha}$ can also be extracted out in our MCMC calculations, which are defined as $g_{ee}(r)=\langle c_1^\dag(\mathbf{r}_1)c_1(\mathbf{r}_1)c_1^\dag(\mathbf{r}_2)c_1(\mathbf{r}_2)\rangle/\nu_1^2$ between two electrons and $g_{eh}(r)=\langle c_1^\dag(\mathbf{r}_1)c_1(\mathbf{r}_1)c_2(\mathbf{r}_2)c_2^\dag(\mathbf{r}_2)\rangle/\nu_1^2$ between an electron and a hole for $r=|\mathbf{r}_1-\mathbf{r}_2|$. The correlations for different values of $\alpha$ are plotted in Fig. \ref{Fig1}(c), where the higher and lower curves are $g_{eh}(r)$ and $g_{ee}(r)$, respectively.
By transforming holes in layer $2$ back to electrons \cite{suppl}, one can show the interlayer electron-electron correlation $g_{12}(r) =\langle c_1^\dag(\mathbf{r}_1)c_1(\mathbf{r}_1)c_2^\dag(\mathbf{r}_2)c_2(\mathbf{r}_2)\rangle/\nu_1\nu_2$
for $r=|\mathbf{r}_1-\mathbf{r}_2|$ is
\begin{equation}
g_{12}(r)=\nu_2^{-1}[1-\nu_1g_{eh}(r)]\ .
\end{equation}
The resulting $g_{12}(r)$ for different $\alpha$ are shown in the inset of Fig. \ref{Fig1}(c). In particular, for $\alpha=1/2$ where both $\nu_1$ and $\nu_2$ are $1/2$, the intralayer and interlayer correlation functions $g_{11}(r)=g_{ee}(r)$ and $g_{12}(r)$ fit remarkably well with previous DMRG results at $d/\ell=\kappa_{c_1}$ \cite{Shibata2006}, as shown in Fig. \ref{Fig2}(a).
This means the energy of the state $\Psi_{1,1/2}$ is quite close to the ground state energy of the system, and thus strongly suggests $\Psi_{1,1/2}$ may be a good approximation to the true ground state at $d/\ell=\kappa_{c1}$. For $\kappa_{c1}=1.1$, the state $\Psi_{1,1/2}$ yields a Coulomb energy per electron $E_0=\sum_{ij}\frac{\nu_i\nu_j}{2\ell^2}\int rV_{ij}(r)[g_{ij}(r)-1]\mbox{d}r\approx -0.35e^2/\epsilon\ell$ \cite{Vij}.

The wave function $\Psi_{1,1/2}$ can be better understood in the picture of CBs \cite{Zhang1989,Read1989,Jain1989,Ye2008}. Here a CB (anti-CB) is defined as an electron (hole) bound with a $2\pi$ statistical flux relative to the other electrons (holes), which obeys bosonic statistics. In the basis of CBs in layer $1$ and anti-CBs in layer $2$, the state $\Psi_{1,1/2}$ can be rewritten as:
\begin{equation}
\Psi^{CB}_{1,1/2}=\mbox{perm}\ \widetilde{M}_{ij}\ ,
\end{equation}
where $\widetilde{M}_{ij}=e^{-(|z_i|^2-2z_iw_{j}^*+|w_j|^2)/8\ell^2}$, while the Jastrow factor in $\Psi_{1,1/2}$ is absorbed by the $2\pi$ fluxes bound to CBs and anti-CBs \cite{Zhang1989,Read1989,Jain1989,Zhang1992}. In analogy to $M_{ij}$ in Eq. (\ref{Psi0}), $\widetilde{M}_{ij}$ is a wave function of a composite exciton formed by the $i$-th CB and the $j$-th anti-CB, where the magnetic field is reduced to $B/2$ due to their bound fluxes. Thus, the state $\Psi^{CB}_{1,1/2}$ can be viewed as a BEC of $N$ \emph{free} composite excitons. This leads us to conjecture that the level crossing point $d/\ell=\kappa_{c1}$ is exactly where the composite excitons become free. An understanding of this is as follows.
As shown in Fig. \ref{Fig1}(f), the interaction between two composite excitons is determined by the intralayer and interlayer interactions $V_{ij}'(q)$ between CBs (anti-CBs).
Due to the fluxes bound to the CBs, the intralayer interaction $V_{11}'(q)=V_{22}'(q)$ is generically largely screened compared to that of electrons \cite{Lee2002}. As a crude estimation, $V_{11}'(q)$ approximately equals to the Coulomb potential between two $e/2$ ($-e/2$) charges, namely, $V_{11}'(q)\approx V_{11}(q)/4$, since the fluxes roughly counteract one half of the gauge potential. In contrast, the interlayer interaction remains $V_{12}'(q)=V_{12}(q)$, since a CB and an anti-CB in different layers do not have a mutual statistical flux. The approximate interaction between two composite excitons is then
\begin{equation}
V_{E}'(q)=V_{11}'+V_{22}'-2V_{12}'\approx2V_{11}(q)\left(\frac{1}{4}-e^{-qd}\right).
\end{equation}
If we substitute $q=1/\ell$ into the formula as a characteristic momentum, we find $V_{E}'(q)$ vanishes at $d/\ell=\ln 4\approx 1.4$, which is rather close to the numerical value $\kappa_{c1}\approx1.1$ considering the roughness of this estimation. A precise determination of $\kappa_{c1}$ would call for a more careful calculation of $V_{ij}'(q)$ in the future.

Similar to the case for electrons at $d=0$, the vanishing of $V_{E}'(q)$ implies an emergent pseudospin SU(2) symmetry for CBs at $d/\ell=\kappa_{c1}$. In condensed matter systems, an emergent symmetry usually leads to extra degeneracies in the energy spectrum \cite{Majumdar1970,Batista2003,Batista2004,Chen2015}. We claim here this emergent SU(2) symmetry is responsible for the level crossing of lowest charged excitations at $d/\ell=\kappa_{c1}$, and should yield approximate degeneracies at higher energies as well. For the same reason, we expect the level crossing to also occur at $d=0$.
This is understood as follows. The minimal charged excitation in the bilayer exciton superfluid is known to be the meron, which has an electrical charge $\pm e/2$ and pseudospin up or down in the core, and evolves into an in-plane pseudospin vortex with vorticity $\pm1$ away from the core \cite{Moon1995,Affleck1986}. In particular, a meron with a given vorticity can have either charge $e/2$ or $-e/2$, and the charge can be localized in either layer depending on the core pseudospin direction \cite{Moon1995,suppl}. A single meron has a logarithmically diverging energy with the system size, thus is not a low energy excitation. However, it is believed that the lowest charged excitation of charge $\pm e$ is a bound state of two merons with the same charge but opposite vorticities \cite{Veillette2002,Ye2008,Moon1995,Moon1998,Milovanovic2015}. There are two kinds of such bound states (BSs) competing: one interlayer BS and two degenerate intralayer BSs (in either layer), where the charges of two merons are in different layers (Fig. \ref{Fig1}(g)) and in the same layer (Fig. \ref{Fig1}(h)), respectively. At $d=0$, these two kinds of BSs are degenerate since they have identical interactions $V_{11}(q)=V_{12}(q)$. Similarly, at $d/\ell=\kappa_{c1}$, we expect the lowest charged excitations to be the interlayer and intralayer bound states of CB merons (Fig. \ref{Fig1}(g) and (h)) \cite{suppl}, which also have identical interactions $V_{11}'(q)\approx V_{12}'(q)$ and are thus degenerate. As a result, we arrive at the low-lying charged excitation spectrum with a level crossing at both $d=0$ and $d/\ell=\kappa_{c1}$ as shown in Fig. \ref{Fig1}(d). In particular, we expect the lower excited state $E_2$ at small $d/\ell$ to adiabatically evolve from the interlayer bound state of merons to that of CB merons, while states $E_1$ and $E_3$ to be the intralayer bound states of merons and CB merons, respectively. We note the level crossing cannot be avoided, since any matrix element between the interlayer BS and the intralayer BS must involve a fractional charge $e/2$ interlayer hopping, which cannot be a local operator and must be zero in the thermodynamic limit. Our results also suggest the intralayer BSs $E_3$ may play a key role in the exciton superfluid-CFL transition at $d/\ell=\kappa_{c2}$.

A disadvantage of the above electron-hole formulation of $\nu_1=\nu_2=1/2$ bilayer is that it is asymmetric between the two layers. As a result, when holes in layer $2$ are transformed back into electrons, the wave function $\Psi_{1,1/2}$ transforms into a bilayer wave function $\Psi_{1/2}$ of $N$ electrons per layer, which is not exactly symmetric between two layers \cite{suppl}. A layer swapping yields its mirror state $\Psi_{1/2}^M=(-1)^N\Psi_{1/2}(z_i\leftrightarrow w_i)$. Numerical calculations show their overlap $|\langle \Psi_{1/2}|\Psi_{1/2}^M\rangle|^2\propto N^{-\alpha}$ with $\alpha\approx 0.5$ \cite{suppl}, so the two states are orthogonal in the thermodynamic limit. However, a power law decay of overlap indicates the two states $\Psi_{1/2}$ and $\Psi_{1/2}^M$ are quite alike each other. In contrast, we find the overlap between $\Psi_{1/2}$ and the Halperin state $\Psi_{111}$ decays exponentially as $|\langle \Psi_{1/2}|\Psi_{111}\rangle|^2\propto e^{-gN^\beta}$ with $\beta\approx 1.5$ \cite{suppl}, indicating they are rather distinct states. Therefore, it is reasonable to believe that $\Psi_{1/2}$ and $\Psi_{1/2}^M$ differ only by some low-energy gapless Goldstone modes \cite{Wen1992b,Yang1994}, and are both close to the ground state at $d/\ell=\kappa_{c1}$. We then propose their symmetric superposition $\Psi_{1/2}^S=\Psi_{1/2}+\Psi_{1/2}^M$ as an improved trial wave function, which respects the mirror symmetry between two layers.

To further test the validity of our theory, we run a small toy size ED calculation for $\nu_1=\nu_2=1/2$ with $N=2$ electrons per layer at different $d/\ell$, and calculate the overlap between the ED ground state $\Psi_{ED}$ and three trial wave functions $\Psi_{111}$, $\Psi_{1/2}$ and $\Psi_{1/2}^S$ in the same total angular momentum sector\cite{suppl}. As shown in Fig. \ref{Fig2}(b), the overlaps of both $\Psi_{1/2}$ and $\Psi_{1/2}^S$ with $\Psi_{ED}$ are indeed peaked at $d/\ell\approx \kappa_{c1}$ as expected in our theory, and the peak value of $|\langle \Psi_{ED}|\Psi_{1/2}^S\rangle|^2$ is as high as $0.95$. In contrast, the overlap between $\Psi_{111}$ and $\Psi_{ED}$ monotonically decays with $d/\ell$.
A larger size ED calculation is desired in the future to further verify this result.

Finally, we mention that wave function $\Psi_{1,\alpha}$ can be generalized into a larger class of wave functions
\begin{equation}
\Psi_{m,\alpha}=\prod_{i<j}^N(z_i-z_j)^m(w_i^*-w_j^*)^m f\left[M_{ij}(\alpha)\right],
\end{equation}
where $f\left[M_{ij}(\alpha)\right]$ is defined as $\det M_{ij}(\alpha)$ for $m$ even, and $\mbox{perm}\ M_{ij}(\alpha)$ for $m$ odd, with $M_{ij}(\alpha)$ defined as in Eq. (\ref{Psi1}). These states admit a similar physical picture of CBs or composite fermions \cite{suppl}, and may describe certain filling-imbalanced $\nu_T=1$ bilayers or electron-hole bilayers \cite{Sivan1992,Sanchez2017}.

In conclusion, we have shown a trial wave function $\Psi_{1,1/2}$ shares many features with the numerical ground state of $\nu_T=1$ quantum Hall bilayer at $d/\ell=\kappa_{c1}=1.1$, and is likely to characterize the ground state well. The wave function implies an emergent SU(2) symmetry for CBs at $d/\ell=\kappa_{c1}$, which gives an straightforward explanation of the excited state level crossing found therein. These results suggest the $\nu_T=1$ bilayer at intermediate $d/\ell$ may have an easier understanding in terms of CBs.


\begin{acknowledgments}
The authors acknowledge helpful conversations with S. A. Kivelson, R. B. Laughlin, S. Raghu and M. P. Zaletel. BL acknowledge the support of Princeton Center for Theoretical Science at Princeton University. SCZ is supported by the NSF grant DMR-1305677.
\end{acknowledgments}

\begin{widetext}
\section{Supplementary Material for "Wave Function and Emergent SU(2) Symmetry in $\nu_T=1$ Quantum Hall Bilayer"}

\subsection{Markov chain Monte Carlo calculations for state $\Psi_{m,\alpha}$}

We employ the Markov chain Monte Carlo (MCMC) method to numerically calculate the electron density in layer 1 $\rho(r)$ (which is also the hole density in layer $2$) and two-particle density-density correlations $g_{ee}(r)$ (between two electrons in layer $1$), $g_{eh}(r)$ (between an electron in layer $1$ and a hole in layer $2$) of the electron-hole trial wave function $\Psi_{1,\alpha}$ defined in Eq. (4) and its generalization $\Psi_{m,\alpha}$ defined in Eq. (8) in the main text. The MCMC method circumvents the difficulty of huge Hilbert space dimension \cite{Girvin1984c}, and is commonly used to obtain the properties of analytical fractional quantum Hall wave functions \cite{Morf1986,Datta1996}. Note that the MCMC numerical calculation here is done in the basis of LLL electrons in layer 1 and LLL holes in layer 2.

To do the MCMC calculation, we regard $p(\{z_i,w_i\})=|\Psi_{m,\alpha}(\{z_i,w_i\})|^2$ as a probability function for $N$ electrons and $N$ holes distributed at coordinates $\{z_i\}$ and $\{w_i\}$ $(1\le i\le N)$, respectively. We then start with a random set of coordinates $\{z_i\}$ and $\{w_i\}$, and move $z_1$ to $z_1'=z_1+u_1$ with a random walk displacement $u_1$ (which obeys a Gaussian distribution). We keep the coordinate change from $z_1$ to $z_1'$ with a probability $\min(p'/p,1)$, where $p=p(\{z_1,z_2,\cdots,w_i\})$ is the initial probability, and $p'=p(\{z_1',z_2,\cdots,w_i\})$ is the probability after the move of $z_1$. Such an operation is then repeated for $z_2$, $z_3$, $\cdots$, $w_1$, $w_2$, $\cdots$, $w_N$, and these $2N$ moves define a step of a Markov chain. After sufficient number of steps, the coordinates of electrons and holes will approach the most probable configurations and achieve an equilibrium. The density $\rho(r)$ is then obtained by averaging the frequency for each electron (hole) coordinate $z_i$ ($w_i$) to be located at position $\mathbf{r}$, and the pair correlations $g_{ee}(r)$, $g_{eh}(r)$ are calculated by averaging the frequency of all pairs of coordinates $(z_i,z_j)$ and $(z_i,w_j)$ with a distance $r$ over all steps which have reached the equilibrium.

In the MCMC calculations for state $\Psi_{1,\alpha}$ in the electron-hole basis, we need to calculate the probability ratio when moving $z_i$ to $z_i'$ (or $w_i$ to $w_i'$)
\begin{equation}
p'/p=\prod_j \left|\frac{z_i'-z_j}{z_i-z_j}\right|^2\left|\frac{\mbox{perm}\ M_{ij}'(\alpha)}{\mbox{perm}\ M_{ij}(\alpha)}\right|^2\ ,
\end{equation}
where $M_{ij}'(\alpha)$ and $M_{ij}(\alpha)$ are defined with $z_i'$ and $z_i$, respectively. However, the permanent part $\mbox{perm}\ M_{ij}(\alpha)$ has an exponential computational cost with respect to the matrix size $N$, which limits the number of electrons (holes) $N$ we can reach in the calculation. We have done the calculations up to system size $N=15$, and find the finite size effects on the pair correlations $g_{ee}(r)$ and $g_{eh}(r)$ are negligible for $N\gtrsim6$. For state $\Psi_{2,\alpha}$ which contains a determinant instead of a permanent, we are able to carry out the MCMC calculations up to $N=20$. The results for $\Psi_{1,\alpha}$ and $\Psi_{2,\alpha}$ are as shown in Fig. 1 in the main text and Fig. \ref{higher} here in the supplementary material, respectively.

\subsection{Two-particle correlations in electron-electron bilayer basis}
The above MCMC calculations are done in the electron-hole basis, namely, electrons in LLL of layer $1$ and holes in LLL of layer $2$, so the two-particle density-density correlation functions $g_{ee}(r)$ and $g_{eh}(r)$ we obtained directly are also in the electron-hole basis. To obtain the two-particle correlations in the electron-electron bilayer basis, one need to do a particle-hole transformation in layer $2$. The particle-hole transformation of the wave function $\Psi_{1,\alpha}$ is quite complicated, as we will discuss later in Sec. \ref{PHtrans} of supplementary material. However, the particle-hole transformation of two-particle correlation functions into the electron-electron basis is much easier, in which one can make use of the fact that the electron (hole) density in each layer is uniform, namely, $\langle c_1^\dag(\mathbf{r})c_1(\mathbf{r})\rangle=\langle c_2(\mathbf{r})c_2^\dag(\mathbf{r})\rangle=\rho(\mathbf{r})=\nu_1$, where $c_i(\mathbf{r})$ and $c_i^\dag(\mathbf{r})$ $(i=1,2)$ are the electron annihilation and creation operators at position $\mathbf{r}$ of layer $i$.

To see this, we first define electron operators $\bar{c}_i(\mathbf{r})$ and $\bar{c}_i^\dag(\mathbf{r})$ which are the LLL projection of $c_i(\mathbf{r})$ and $c_i^\dag(\mathbf{r})$.
More specifically, $\bar{c}_i^\dag(\mathbf{r})$ creates the coherent LLL electron state $e^{-(|z|^2+2\xi_\mathbf{r}^*z-|\xi_\mathbf{r}|^2)/4\ell^2}$ with guiding center at $\mathbf{r}$, where $\xi_\mathbf{r}=x_\mathbf{r}+iy_\mathbf{r}$ is the complex coordinate of $\mathbf{r}=(x_\mathbf{r},y_\mathbf{r})$, and $z$ is the complex electron coordinate. A straightforward calculation \cite{Bargmann1962} yields the anti-commutation relation $\{\bar{c}_i(\mathbf{r}_1),\bar{c}_j^\dag(\mathbf{r}_2)\}=\delta_{ij}e^{-(|\xi_{\mathbf{r}_1}|^2 -2\xi_{\mathbf{r}_1}\xi_{\mathbf{r}_2}^*+|\xi_{\mathbf{r}_2}|^2)/4\ell^2}$, where we have set the density of a fully occupied Landau level to $1$. In particular, this shows $\{\bar{c}_i(\mathbf{r}),\bar{c}_j^\dag(\mathbf{r})\}=\delta_{ij}$.

The density-density correlation between two electrons is defined by $g_{ij}(r)=\langle c_i^\dag(\mathbf{r}_1) c_i(\mathbf{r}_1) c_j^\dag(\mathbf{r}_2) c_j(\mathbf{r}_2)\rangle/\nu_i\nu_j$ in terms of projected fermion operators, where $r=|\mathbf{r}_1-\mathbf{r}_2|$, and $\nu_i$ is the filling fraction of layer $i$. In comparison, the density-density correlations in the electron-hole basis are given by $g_{ee}(r)=\langle c_1^\dag(\mathbf{r}_1) c_1(\mathbf{r}_1) c_1^\dag(\mathbf{r}_2) c_1(\mathbf{r}_2)\rangle/\nu_1^2$ and $g_{eh}(r)=\langle c_1^\dag(\mathbf{r}_1) c_1(\mathbf{r}_1) c_2(\mathbf{r}_2) c_2^\dag(\mathbf{r}_2)\rangle/\nu_1(1-\nu_2)$. The definitions already show $g_{11}(r)=g_{ee}(r)$. To see the relation between $g_{12}(r)$ and $g_{eh}(r)$, we note the fact that $P_{LLL}^ic_i^\dag(\mathbf{r})c_i(\mathbf{r})P_{LLL}^i=\bar{c}_i^\dag(\mathbf{r})\bar{c}_i(\mathbf{r})$ and $P_{LLL}^ic_i(\mathbf{r})c_i^\dag(\mathbf{r})P_{LLL}^i=\bar{c}_i(\mathbf{r})\bar{c}_i^\dag(\mathbf{r})$, where $P_{LLL}^i$ is the projector into the LLL of layer $i$. Then, using the fact that the trial wave function is in the LLLs of bilayer, and the identities $\{\bar{c}_2(\mathbf{r}),\bar{c}_2^\dag(\mathbf{r})\}=1$ and $\nu_1+\nu_2=1$, one finds
\begin{equation}
\begin{split}
&\nu_1^2g_{eh}(r)+\nu_1\nu_2g_{12}(r)=\langle c_1^\dag(\mathbf{r}_1) c_1(\mathbf{r}_1) c_2(\mathbf{r}_2) c_2^\dag(\mathbf{r}_2)\rangle+\langle c_1^\dag(\mathbf{r}_1) c_1(\mathbf{r}_1) c_2^\dag(\mathbf{r}_2) c_2(\mathbf{r}_2)\rangle \\
&=\langle P_{LLL}^1 P_{LLL}^2c_1^\dag(\mathbf{r}_1) c_1(\mathbf{r}_1) c_2(\mathbf{r}_2) c_2^\dag(\mathbf{r}_2) P_{LLL}^1P_{LLL}^2\rangle+\langle P_{LLL}^1P_{LLL}^2 c_1^\dag(\mathbf{r}_1) c_1(\mathbf{r}_1) c_2^\dag(\mathbf{r}_2) c_2(\mathbf{r}_2) P_{LLL}^1P_{LLL}^2\rangle\\
&=\langle P_{LLL}^1 c_1^\dag(\mathbf{r}_1) c_1(\mathbf{r}_1)P_{LLL}^1 P_{LLL}^2c_2(\mathbf{r}_2) c_2^\dag(\mathbf{r}_2) P_{LLL}^2\rangle+\langle P_{LLL}^1 c_1^\dag(\mathbf{r}_1) c_1(\mathbf{r}_1)P_{LLL}^1 P_{LLL}^2 c_2^\dag(\mathbf{r}_2) c_2(\mathbf{r}_2) P_{LLL}^2\rangle\\
&=\langle \bar{c}_1^\dag(\mathbf{r}_1) \bar{c}_1(\mathbf{r}_1) \bar{c}_2(\mathbf{r}_2) \bar{c}_2^\dag(\mathbf{r}_2)\rangle+\langle \bar{c}_1^\dag(\mathbf{r}_1) \bar{c}_1(\mathbf{r}_1) \bar{c}_2^\dag(\mathbf{r}_2) \bar{c}_2(\mathbf{r}_2)\rangle\\
&=\langle \bar{c}_1^\dag(\mathbf{r}_1) \bar{c}_1(\mathbf{r}_1)\rangle=\nu_1\ ,
\end{split}
\end{equation}
from which one derives the relation $g_{12}(r)=\nu_2^{-1}[1-\nu_1g_{eh}(r)]$.

Another quantity one could extract from the electron-hole wave function is the hole-hole correlation $g_{hh}(r)=\langle c_2(\mathbf{r}_1) c_2^\dag(\mathbf{r}_1) c_2(\mathbf{r}_2) c_2^\dag(\mathbf{r}_2)\rangle/(1-\nu_2)^2$ in layer $2$, which is equal to $g_{ee}(r)$. A similar but more complicated analysis shows the electron-electron correlation $g_{22}(r)$ in layer $2$ satisfies approximately $g_{22}(r)=\nu_2^{-2}[(2\nu_2-1)g_{L}(r)+(1-\nu_2)^2g_{hh}(r)]$. A heuristic way to see this is the following: if we simply replace $c_2(\mathbf{r})$ by its LLL projection $\bar{c}_2(\mathbf{r})$, we would have
\begin{equation}
\begin{split}
\langle \bar{c}_2(\mathbf{r}_1) \bar{c}_2^\dag(\mathbf{r}_1) \bar{c}_2(\mathbf{r}_2) \bar{c}_2^\dag(\mathbf{r}_2)\rangle= \langle (1-\bar{c}_2^\dag(\mathbf{r}_1) \bar{c}_2(\mathbf{r}_1)) (1-\bar{c}_2^\dag(\mathbf{r}_2) \bar{c}_2(\mathbf{r}_2))\rangle=(1-2\nu_2)+\langle \bar{c}_2^\dag(\mathbf{r}_1) \bar{c}_2(\mathbf{r}_1) \bar{c}_2^\dag(\mathbf{r}_2) \bar{c}_2(\mathbf{r}_2)\rangle\ .
\end{split}
\end{equation}
Such a replacement, however, ignores the correlations between two electrons in a fully occupied LLL. This is corrected by modifying the correlations between two "$1$" into the two-particle correlation of a fully occupied LLL $g_L(r)=1-e^{-r^2/2\ell^2}$. After this modification, we arrive at the approximate relation $g_{22}(r)=\nu_2^{-2}[(2\nu_2-1)g_{L}(r)+(1-\nu_2)^2g_{hh}(r)]$. In particular, in the case $\nu_1=\nu_2=1/2$ which we are most interested in, we have $g_{22}(r)=g_{hh}(r)=g_{ee}(r)=g_{11}(r)$.

\subsection{Merons in the $\nu_T=1$ quantum Hall bilayer}

A symmetric $\nu_T=1$ bilayer in an exciton superfluid state is known to have a pseudospin polarized in the $x$-$y$ plane. A meron in such a system is a topologically stable pseudospin configuration that points up or down in the core, while lies in the $x$-$y$ plane forming a vortex with vorticity $\pm1$ far away from the core. Therefore, it can be viewed as a half-skyrmion. It has been shown that a meron necessarily carries a half charge $\pm e/2$ in the meron core \cite{Moon1995}.

In the case $d\rightarrow0$ where the ground state is the Halperin (111) state, a meron can be conveniently expressed in the second quantized representation of electrons. We can define $\bar{c}^\dag_{\uparrow,l}$ and $\bar{c}^\dag_{\downarrow,l}$ as the electron creation operator of the angular momentum $l$ LLL state $z^{l}e^{-|z|^2/4\ell^2}/\sqrt{2^ll!}$ in layer $1$ (pseudospin up) and layer $2$ (pseudospin down), respectively. In this representation, the ground state as $d\rightarrow 0$ is a pseudospin polarized state
\begin{equation}
|+x\rangle=\prod_{l=0}^{N_{L}}\frac{\bar{c}^\dag_{\uparrow,l}+\bar{c}^\dag_{\downarrow,l}}{\sqrt{2}}|0\rangle\ ,
\end{equation}
where $N_L=N+M$ is the degeneracy of a single Landau level, and we have assumed the polarization is along $+x$ direction. We note that in this representation, the numbers of electrons $N$ in layer 1 and $M$ in layer 2 are indefinite, but are both peaked at $N_L/2$. If one projects the state $|+x\rangle$ into a subspace with definite $N$ and $M$, the state becomes the Halperin state $\Psi_{111}$.

If we denote a meron with vorticity $V=\pm1$, charge $Q=\pm e/2$ and pseudospin direction $s=\uparrow,\downarrow$ in the meron core as $|m_{V,Q,s}\rangle$, we could write down a meron with any combinations of the three topological indices \cite{Moon1995}:
\begin{equation}
\begin{split}
|m_{+1,-\frac{e}{2},s}\rangle=\frac{\bar{c}^\dag_{s,0}+\bar{c}^\dag_{\downarrow,1}}{\sqrt{2}} \prod_{l=1}^{N_L}&\frac{\bar{c}^\dag_{\uparrow,l}+\bar{c}^\dag_{\downarrow,l+1}}{\sqrt{2}}|0\rangle, \qquad |m_{-1,-\frac{e}{2},s}\rangle=\frac{\bar{c}^\dag_{\uparrow,1}+\bar{c}^\dag_{s,0}}{\sqrt{2}} \prod_{l=1}^{N_L}\frac{\bar{c}^\dag_{\uparrow,l+1}+\bar{c}^\dag_{\downarrow,l}}{\sqrt{2}}|0\rangle,\\ &\ \ |m_{V,+\frac{e}{2},s}\rangle=\bar{c}^\dag_{s,0}|m_{V,-\frac{e}{2},-s}\rangle.
\end{split}
\end{equation}
The pseudospin direction $s$ determines the layer the charge $Q$ is located in. A positive charge $Q=e/2$ is located in layer $s$, while a negative charge $Q=-e/2$ is located in layer $-s$. It should be mentioned that these are the merons with minimal core sizes one could write down. While the core sizes may vary in realistic systems to lower the energy, the topological characters of the merons remain. Similarly, the meron states can also be projected into the subspace with definite numbers of particles $N$ and $M$ in layers 1 and 2. In particular, after such a procedure, the meron states $|m_{+1,-\frac{e}{2},\downarrow}\rangle$ and $|m_{+1,-\frac{e}{2},\uparrow}\rangle$ are topologically equivalent to the quasi-hole wave function in layer 1 and layer 2 centered at the origin given by $\left(\prod_{i=1}^{N}z_i\right)\Psi_{111}$ and $\left(\prod_{i=1}^{N}w_i\right)\Psi_{111}$ \cite{Jeon2005}, respectively, which are known to carry fractional charges $-e/2$.

\begin{figure}[htpb]
\begin{center}
\includegraphics[width=4in]{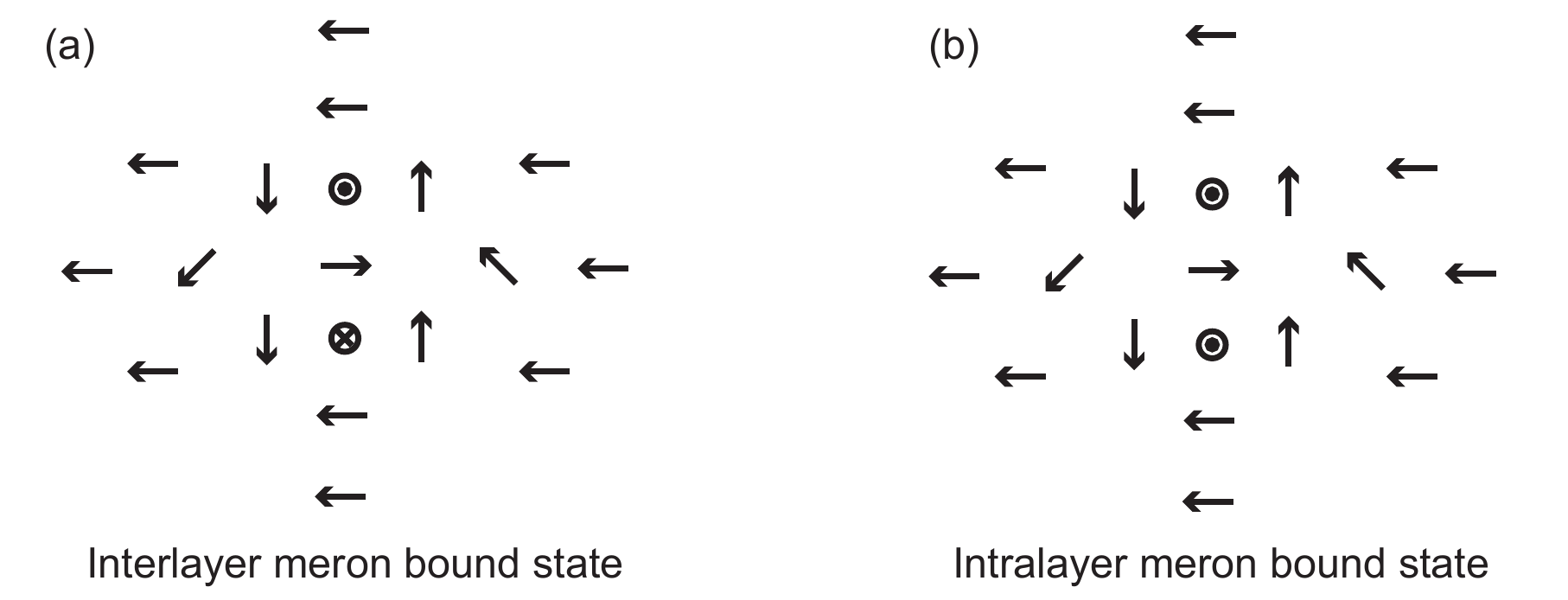}
\end{center}
\caption{Top view of the spin configurations of interlayer (a) and intralayer meron bound states carrying charge $-e$. In the plot $\odot$ and $\otimes$ stand for arrows pointing outside and into the paper, respectively.}
\label{Meron}
\end{figure}

Like vortices in the XY model, the energy of a meron is logarithmically divergent with the system size \cite{Moon1995}, therefore is not a low energy excitation. In contrast, a pair of merons with opposite vorticities (a vortex anti-vortex pair) but the same charge can form a charged bound state which has a finite energy \cite{Moon1995}. Besides, depending on whether the core charges of the two merons are in the same layer or different layers, one can have either an intralayer bound state or an interlayer bound state. For instance, in the $d\rightarrow0$ case, two merons $|m_{+1,-\frac{e}{2},s}\rangle$ and $|m_{-1,-\frac{e}{2},s'}\rangle$ at a distance $r$ will have a vorticity-induced logarithmic attraction potential $V_m(r)\propto \ln(r/r_{mc})$ where $r_{mc}$ is the meron core size, and a Coulomb repulsion due to the half-charges $-e/2$ at the meron cores, which is $V_{11}/4$ intralayer (for $s=s'$) and $V_{12}/4$ interlayer (for $s=-s'$). Therefore, the total potential has a minimum at a finite distance $r_b$, which is the distance between two merons in the bound state. In particular, as shown in the top view plotted in Fig. \ref{Meron}, the pseudospin configuration of the interlayer bound state is exactly a skyrmion, while that of the intralayer bound state has a topologically trivial pseudospin configuration. As $d\rightarrow0$ one has $V_{11}=V_{12}$, so the two kinds of bound states will become degenerate.

Similarly, in the case the ground state is $\Psi_{1,\frac{1}{2}}$ in the electron-hole representation, one can define the layer indices of the CBs as a pseudospin, so the ground state still has an in-plane polarized pseudospin, which is exactly $\Psi_{1,\frac{1}{2}}$ when projected into the subspace with definite number of electrons (holes) $N$. Meron states $|m_{V,Q,s}'\rangle$ with vorticity $V=\pm1$, charge $\pm e/2$ and central pseudospin direction $s=\uparrow,\downarrow$ can be written down for CBs in a similar way as above. Most importantly, these merons are still charged $\pm e/2$, which is solely determined by the filling fraction of charges $\nu_1=\nu_2=1/2$ in each of the two layers and is topologically protected \cite{Arovas1984,Moon1995}. In particular, as an example, the meron state $|m_{+1,-e/2,\downarrow}'\rangle$ after a projection into the supspace of definite number of electrons (holes) $N$ is topologically equivalent to the quasi-hole state $\left(\prod_{i=1}^{N}z_i\right)\Psi_{1,\frac{1}{2}}$ centered at the origin, which can be easily shown to carry charge $-e/2$ via the standard argument \cite{Arovas1984}.

Similarly, the merons of CBs have both interlayer and intralayer bound states, except that they have Coulomb repulsions between merons replaced by $V_{11}'/4$ and $V_{12}'/4$. Therefore, when $d/\ell=\kappa_{c1}$ where we predict $V_{11}'=V_{12}'$, the two kinds of bound states become degenerate.

\subsection{Particle-Hole transformation of $\Psi_{1,1/2}$}\label{PHtrans}
We have been using the electron-hole picture the $\nu_T=1$ electron-electron bilayer to formulate our trial wave function $\Psi_{1,1/2}$ for ground state at $d/\ell=\kappa_{c1}$, which yields an intuitive physical picture of CB exciton condensation. In practice, it is also useful to obtain its form in the electron-electron basis of the bilayer by doing a particle-hole transformation within the LLL of layer $2$.

As shown by Ref. \cite{Girvin1984a}, the particle-hole transformation of an electron wave function $\Psi(z_1,\cdots,z_N)$ within the LLL leads it into a hole wave function
\begin{equation}\label{eh}
\Psi_h(z_{N+1}^*,\cdots,z_{N_L}^*)=\int \prod_{i=1}^N dz_i^*dz_i \left[\prod_{i=1}^{N_L}e^{-|z_i|^2/4\ell^2}\prod_{1=i<j}^{N_L}(z_i^*-z_j^*)\right]\Psi(z_1,\cdots,z_N)\ ,
\end{equation}
where $N_L$ is the Landau level degeneracy. Based on this, it has been shown that the Halperin state $\Psi_{111}$ and $\Psi_0$ in Eqs. (1) and (2) of the main text transform into each other after a particle-hole transformation in layer $2$ \cite{Yang2001}.

Here we present the transformation of our electron-hole trial wave function $\Psi_{1,\alpha}(z_1,\cdots,z_N,w_1^*,\cdots,w_N^*)$ into electron-electron basis through a particle-hole transformation in layer $2$. According to Eq. (\ref{eh}) above, the resulting electron-electron wave function is
\begin{equation}\label{P12}
\begin{split}
&\Psi_{\alpha}(z_1,\cdots,z_N,w_{N+1},\cdots,w_{N_L})= \int \prod_{i=1}^N dw_i^*dw_i \left[\prod_{i=1}^{N_L}e^{-|w_i|^2/4\ell^2}\prod_{1=i<j}^{N_L}(w_i-w_j)\right] \Psi_{1,\alpha}(z_1,\cdots,w_N^*)\\
&=\int \prod_{i=N+1}^{N_L}e^{-|w_i|^2/4\ell^2}\prod_{1=i<j}^{N_L}(w_i-w_j) \left[\prod_{1=i<j}^{N}(z_i-z_j)(w_i^*-w_j^*)\right]\left[\sum_\sigma \prod_{i=1}^N dw_i^*dw_i e^{-(|z_{\sigma_i}|^2/2-\alpha z_{\sigma_i}w_i^*+|w_i|^2)/2\ell^2}\right]\\
&=\mu(z,w) \left[\prod_{1=i<j}^{N}(z_i-z_j)(\partial_{z_i}-\partial_{z_j})\right]\left[\sum_\sigma \int \prod_{i=1}^N dw_i^*dw_i e^{-(|w_i|^2-\alpha z_{\sigma_i}w_i^*)/2\ell^2}\prod_{1=i<j}^{N_L}(w_i-w_j)\right]\\
&=\mu(z,w) \left[\prod_{1=i<j}^{N}(z_i-z_j)(\partial_{z_i}-\partial_{z_j})\right]\left[\sum_\sigma \int d^2w_i\prod_{i=1}^N \delta^2( \alpha z_{\sigma_i}-w_i)\prod_{1=i<j}^{N_L}(w_i-w_j)\right]\\
&=\mathcal{N}^{-1}\mu(z,w)\prod_{1=i<j}^{N}(z_i-z_j)\prod_{N+1=i<j}^{N_L}(w_i-w_j) \prod_{1=i<j}^{N}(\partial_{z_i}-\partial_{z_j}) \left[\prod_{1=i<j}^{N}(z_i-z_j) \prod_{i=1}^{N}\prod_{j=N+1}^{N_L}(z_i-\alpha^{-1}w_j)\right]\ ,
\end{split}
\end{equation}
where $\mu(z,w)=\prod_{i=1}^{N}e^{-|z_i|^2/4\ell^2}\prod_{i=N+1}^{N_L}e^{-|w_i|^2/4\ell^2}$ is the Gaussian factor, $\mathcal{N}$ is a numerical factor, and we have used the Bargmann identity $\int dz^*dz f(w)e^{-|z|^2+w^*z}=f(z)$ in Ref. \cite{Bargmann1962}.

\begin{figure}[tbp]
\begin{center}
\includegraphics[width=4.5in]{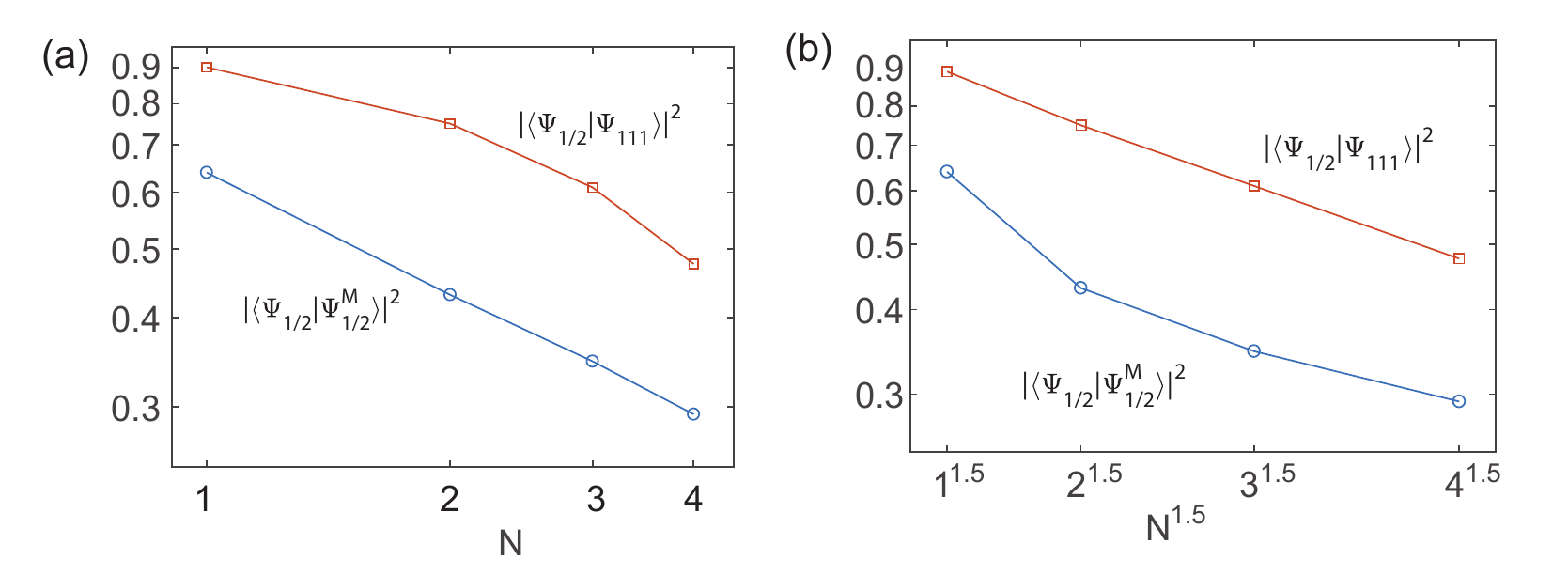}
\end{center}
\caption{(color online) (a) Log-Log plot of overlaps $|\langle\Psi_{1/2}|\Psi_{1/2}^M\rangle|^2$ and $|\langle\Psi_{1/2}|\Psi_{111}\rangle|^2$ as a function of $N$. The horizontal axis is $\mbox{Log} N$, while the vertical axis is linear to Log of the overlap. This plot shows $|\langle\Psi_{1/2}|\Psi_{1/2}^M\rangle|^2\propto N^{-\alpha}$, with $\alpha\approx 0.5$, while $|\langle\Psi_{1/2}|\Psi_{111}\rangle|^2$ does not follow a power law of $N$. (b) Log of the overlaps $|\langle\Psi_{1/2}|\Psi_{1/2}^M\rangle|^2$ and $|\langle\Psi_{1/2}|\Psi_{111}\rangle|^2$ (vertical axis) versus $N^{1.5}$ (horizontal axis), from which we conclude $\mbox{Log}|\langle\Psi_{1/2}|\Psi_{111}\rangle|^2 \propto N^{1.5}$ holds approximately.}
\label{scaling}
\end{figure}

At $\alpha=1/2$, the form of the above wave function $\Psi_{1/2}$ is certainly not symmetric about the two layers. This is because the electron-hole picture we start from is not symmetric between two layers. Besides, the factor $\prod(z_i-\alpha^{-1}w_j)=\prod(z_i-2w_j)$ in the expression of Eq. (\ref{P12}) looks like a nonlocal correlation. However, because of the $\prod (\partial_{z_i}-\partial_{z_j})$ factor in front of it, it is hard to draw such a conclusion. As a simple example, one can easily rewrite $(z_1-w_1)$ as $\partial_{z_1}[z_1(z_1-2w_1)]$, while nothing is nonlocal there. Since nothing is nonlocal in the electron-hole picture, we believe there is nothing nonlocal in the full expression of the electron-electron wave function.

A simple way to recover the layer swap symmetry is to define its mirror state $\Psi_{1/2}^M=(-1)^N\Psi_{1/2}(z_i\leftrightarrow w_i)$ (where $(-1)^N$ is to cancel the sign of the mutual factor $\prod (z_i-2w_j)$), and define the symmetric state $\Psi_{1/2}^S=\Psi_{1/2}+\Psi_{1/2}^M$ as a trial wave function that respects the layer swap symmetry. As we discussed in the main text, we expect $\Psi_{1/2}$ and $\Psi_{1/2}^M$ to only differ by certain creation and annihilation of low energy Goldstone modes, and are both close to the ground state. To see this, we have computed the overlap between $\Psi_{1/2}$ and $\Psi_{1/2}^M$ as a function of $N$. Due to the exponential computational complexity of the above wave function $\Psi_{1/2}$ in electron-electron basis, we are only able to carry out the calculations up to $N=4$, which, however, already shows useful information. As shown in Fig. \ref{scaling}(a), we find the overlap between the two states obeys a power law $|\langle\Psi_{1/2}|\Psi_{1/2}^M\rangle|^2\propto N^{-\alpha}$, with the exponent $\alpha\approx 0.5$. Such a power law decay indicates the two states are quite close to each other, usually differ only by some low energy gapless excitations, though they are orthogonal as $N\rightarrow\infty$. In this system, these gapless excitations are very likely to be the Goldstone modes. In contrast, calculation shows the overlap between $\Psi_{1/2}$ and the Halperin state $\Psi_{111}$ decays much faster. As shown in Fig. \ref{scaling}(b), we find $|\langle\Psi_{1/2}|\Psi_{111}\rangle|^2\propto e^{-gN^{\beta}}$, where $g>0$ is a constant, and $\beta\approx 1.5$. This clearly indicates state $\Psi_{1/2}$ is completely different from the Halperin (111) state.

\subsection{Small Size ED calculation}
We carried out a small size ED calculation for a bilayer at $\nu_1=\nu_2=1/2$ with $N=2$ particles per layer, and calculate the overlap between the ED ground state $\Psi_{ED}$ and trial wave functions $\Psi_{1/2}$, $\Psi_{1/2}^M$ and $\Psi_{111}$. For simplicity, the ED is carried out on a disk instead of a closed manifold, and we restrict the electrons in each layer to occupy Landau level states of angular momentum $0\le l\le3$ only. For $N=2$, this corresponds to half filling in each layer. The Coulomb interactions $V_{11}(r)=e^2/r$ and $V_{12}(r)=e^2/\sqrt{r^2+d^2}$ are then projected into this Hilbert subspace, which forms the Hamiltonian for ED.

A subtlety of this ED calculation is that, since the disk has a boundary, the ground state energies may suffer from the effect of both gapless edge states and finite system size, leading to some other states to go below the true ground state in the thermodynamic limit. In particular, a gapless edge state excitation will lead to a state with different total angular momentum but similar energy. To circumvent this difficulty, we restrict the total angular momentum of the system to be the same as that of the three trial wave functions (which are all equal), and identify the lowest energy state as the ground state $\Psi_{ED}$. The resulting overlaps as a function of $d/\ell$ is shown in Fig. 2(b) of the main text. In particular, the symmetrized trial state $\Psi_{1/2}^S$ has the highest overlap with $\Psi_{ED}$ at $d/\ell\approx\kappa_{c1}$.

\subsection{Physical picture of generalized states $\Psi_{m,\alpha}$}

In Eq. (8) of the main text we wrote down a generalized wave function $\Psi_{m,\alpha}$ in the electron-hole basis. In particular, $\Psi_{0,1}$ is exactly the exciton wave function $\Psi_0$ in Eq. (2) of the main text for the Halperin state. Similarly, wave function $\Psi_{m,\alpha}$ can be understood as a BEC of free composite excitons of higher sequence CFs or CBs and their antiparticles (Fig. \ref{higher}(d)), which are defined as electrons (holes) bound with $2m\pi$ fluxes. As an example, we examine here the properties of state $\Psi_{2,\alpha}$. Fig. \ref{higher} (a)-(c) shows the filling factor $\nu_1$, density profile $\rho(r)$ and two-particle correlations $g_{ee}(r)$ and $g_{eh}(r)$ of $\Psi_{2,\alpha}$ calculated from MCMC, respectively. In particular, the filling factor is found to be $\nu_1=(1-\alpha)/2$ when $\alpha>1/3$, and $\nu_1=1/3$ when $\alpha\le1/3$. The filling $\nu_1=(1-\alpha)/2$ when $\alpha>1/3$ can be understood as equal to that of a $\nu=1/m$ Laughlin state in a magnetic field $(1-\alpha)B$ following the argument for $\Psi_{1,\alpha}$. However, this picture breaks down when $\nu_1$ reaches $1/3$. In the perspective of CFs bound with $4\pi$ fluxes as shown in Fig. \ref{higher}(e), this is because the LLLs of the CFs and anti-CFs in the bilayer are fully occupied when $\alpha\le1/3$, so the state becomes two copies of $\nu_1=1/3$ Laughlin states. In general, as verified by the MCMC calculation, the state $\Psi_{m,\alpha}$ has $\nu_1=(1-\alpha)/m$ for all $0\le\alpha\le1$ if $m$ is odd; while the state has $\nu_1=(1-\alpha)/m$ for $\alpha>1/(m+1)$ and $\nu_1=1/(m+1)$ for $\alpha\le1/(m+1)$ if $m$ is even.

This distinction between even $m$ and odd $m$ is exactly due to different statistics obeyed by CFs and CBs. Since $\nu_1$ for $m>1$ is always smaller than $1/2$, such states may only occur in an asymmetric $\nu_T=1$ bilayer system with proper interactions. It is also possible to realize these states in an electron-hole bilayer with equal number of electrons and holes, which is achievable in GaAs/AlGaAs and graphene systems \cite{Sivan1992,Sanchez2017}.

\begin{figure}[tbp]
\begin{center}
\includegraphics[width=3.4in]{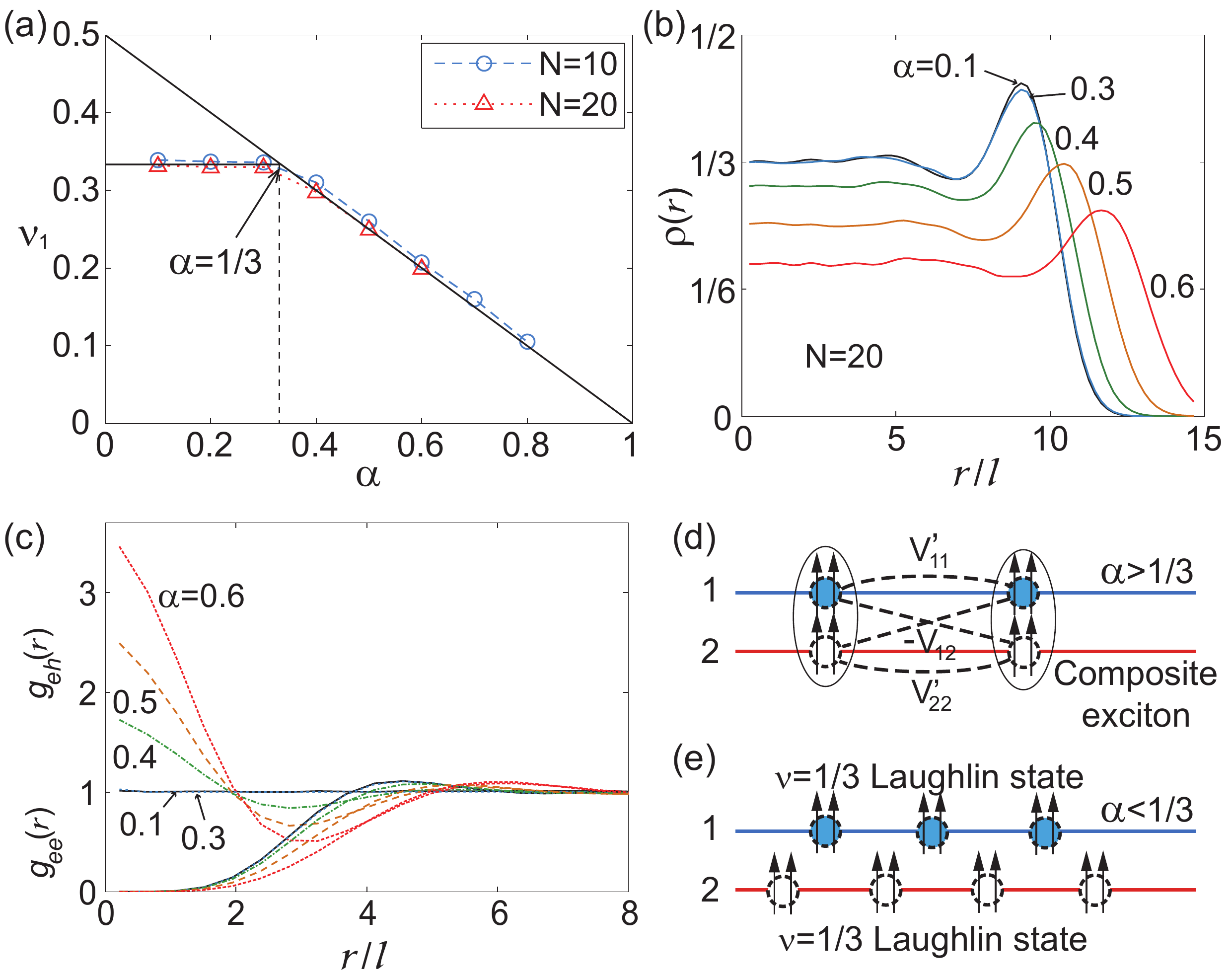}
\end{center}
\caption{(color online) (a) The filling factor $\nu_1$ of state $\Psi_{2,\alpha}$ as a function of $\alpha$ from MCMC. (b) The density profile $\rho(r)$ of $\Psi_{2,\alpha}$ of size $N=20$ for $\alpha=0.1\sim0.6$. (c) The two-particle correlations $g_{ee}(r)$ and $g_{eh}(r)$ for different $\alpha$. (d) When $\alpha>1/3$, $\Psi_{2,\alpha}$ is a BEC of composite excitons of CFs. (e) When $\alpha\le1/3$, the LLL of CFs is fully occupied, and $\Psi_{2,\alpha}$ becomes two copies of the $1/3$ Laughlin state.}
\label{higher}
\end{figure}

In the below we explain analytically why the filling factor of $\Psi_{m,\alpha}$ with an even $m$ is exactly $\nu_1=1/(m+1)$ when $\alpha$ is small enough. The key observation is that the determinant $\det M_{ij}(\alpha)$ can be expanded as
\begin{equation}\label{expand}
\begin{split}
\det &M_{ij}(\alpha)=\mu(z,w)\det\left(e^{\alpha z_iw_j^*/2}\right)=\mu(z,w)\det \left(\sum_{n=0}^\infty\frac{\alpha^n z_i^nw_j^{*n}}{2^n n!}\right)= \mu(z,w) \sum_{\sigma}\mbox{sgn}(\sigma)\prod_{i=1}^N \left(\sum_{n=0}^\infty\frac{\alpha^n z_i^nw_{\sigma_i}^{*n}}{2^n n!}\right)\\
&=\mu(z,w) \left[\frac{\alpha^{N(N-1)/2}}{2^{N(N-1)/2}0!1!\cdots n!}\left(\sum_{\sigma}\mbox{sgn}(\sigma)\prod_{i=1}^N z_i^{\sigma_i}\right)\left(\sum_{\sigma'}\mbox{sgn}(\sigma')\prod_{i=1}^N w_i^{*\sigma_i'}\right)+\mathcal{O}\left(\alpha^{N(N-1)/2+1}\right) \right]\\
&=\mu(z,w) \left[\frac{\alpha^{N(N-1)/2}}{2^{N(N-1)/2}0!1!\cdots n!}\prod_{i<j}^N(z_i-z_j)(w_i-w_j)+\mathcal{O}\left(\alpha^{N(N-1)/2+1}\right) \right]\ .
\end{split}
\end{equation}
Therefore, when $\alpha$ is sufficiently small so that we can keep only the leading order of $\alpha$, we will find
\begin{equation}
\Psi_{m,\alpha}=A\mu(z,w)\prod_{i<j}^N(z_i-z_j)^{m+1}(w_i-w_j)^{m+1}
\end{equation}
for even $m$, where $A$ is a constant factor. Namely, $\Psi_{m,\alpha}$ with even $m$ and a small enough $\alpha$ is simply equivalent to two copies of $1/(m+1)$ Langhlin states for electrons and holes, respectively, in agreement with the MCMC results. This picture, however, becomes invalid when $\alpha$ is large so that the higher order terms in $\alpha$ cannot be ignored.

A special case is when $m=0$, where we find the filling factor $\nu_1=1$ for any $0<\alpha<1$. On the contrary, when $\alpha=1$, the wave function $\Psi_{0,1}$ is exactly $\Psi_0$ defined in Eq. (2) of the main text and can have an arbitrary $\nu_1$. In fact, when $\alpha=1$, the wave function $\Psi_{0,1}$ is no longer bounded in a finite radius $r$. To ensure the wave function make sense, one has to restrict the wave function inside a finite radius $r<R=\sqrt{2N_L}\ell$. Via a derivation similar to Eq. (\ref{expand}) in the above one can show
\begin{equation}
\Psi_{0,1}=\det M_{ij}(1)=\sum_{\{l_i\}\in\{0,1,2,\cdots,N_L-1\}}\prod_{i=1}^{N}c_{\uparrow,l_i}^\dag h_{\downarrow,l_i}^\dag|0\rangle\ ,
\end{equation}
where $c_{\uparrow,l}^\dag$ and $h_{\downarrow,l}^\dag$ are the creation operators of the angular momentum $l$ electron state $z^{l}e^{-|z|^2/4\ell^2}/\sqrt{2^ll!}$ in layer $1$ and hole state $w^{*l}e^{-|w|^2/4\ell^2}/\sqrt{2^ll!}$ in layer $2$, respectively. By controlling the value of $R$, the filling factor $\nu_1=N/N_L$ can take any value between $0$ and $1$.

\end{widetext}


\begin{thebibliography}{69}
\bibitem[{per()}]{perm}
\bibinfo{note}{The permanent of a matrix $M$ is defined as $\mbox{perm}\
  M_{ij}=\sum_{\sigma}\prod_{i}M_{i\sigma_i}$, where $\sigma$ runs over all
  permutations.}
\bibitem[{Vij()}]{Vij}
\bibinfo{note}{Here $V_{11}(r)=e^2/\epsilon r$ and
  $V_{12}(r)=e^2/\epsilon\sqrt{r^2+d^2}$ are unprojected interactions.}
\end{thebibliography}

\end{document}